\documentclass[aps,prd,twocolumn,preprintnumbers, groupedaddress,nofootinbib,amssymb,notitlepage,eqsecnum]{revtex4-2}
\usepackage{bm}
\usepackage{amsmath,amsthm,amssymb}
\usepackage{amsfonts}
\usepackage{hyperref}
\usepackage[dvipdfmx]{graphicx}

\allowdisplaybreaks[1]

\newcommand{\rd}{{\rm d}}
\newcommand{\be}{\begin{equation}}
\newcommand{\ee}{\end{equation}}
\newcommand{\ba}{\begin{eqnarray}}
\newcommand{\ea}{\end{eqnarray}}

\begin{document}

\preprint{WUCG-25-06}

\title{Strong coupling and instabilities in singularity-free inflation from \\
an infinite sum of curvature corrections}

\author{Shinji Tsujikawa}

\affiliation{
Department of Physics, Waseda University, 3-4-1 Okubo, 
Shinjuku, Tokyo 169-8555, Japan}
 
\date{\today}

\begin{abstract}

Four-dimensional gravitational theories derived from an infinite sum of Lovelock curvature invariants, combined with a conformal rescaling of the metric, are equivalent to a subclass of shift-symmetric Horndeski theories that possess a single scalar degree of freedom. 
Under the assumption of a homogeneous and isotropic cosmological background, the theory admits an inflationary solution that replaces the Big Bang singularity. 
This can be achieved by a solution where the Hubble expansion rate $H$ is equal to the time derivative of the scalar field $\dot{\phi}$. We show that the solution $H=\dot{\phi}$ suffers from a strong coupling problem, characterized by the vanishing kinetic term 
of linear scalar perturbations at all times. 
Consequently, nonlinear scalar perturbations remain uncontrolled from the onset of inflation throughout the subsequent cosmological evolution.
Moreover, tensor perturbations are generally subject to Laplacian instabilities during inflation. 
This instability in the tensor sector also persists under background initial conditions where $H \neq \dot{\phi}$. 
In the latter case, both the coefficient of the kinetic term for scalar perturbations and the scalar sound speed diverge at the onset of inflation.
Thus, the dominance of inhomogeneities in this theory renders the homogeneous background solution illegitimate.

\end{abstract}

%\pacs{04.50.Kd,95.30.Sf,98.80.-k}

\maketitle

%%%%%%%%%%%%%%%%%%%%%%%%%%%%%
\section{Introduction}
\label{Intro}
%%%%%%%%%%%%%%%%%%%%%%%%%%%%%

Under several physically reasonable assumptions about spacetime and matter, the appearance of singularities at the centers of black holes (BHs) or at the onset of the Big Bang is a natural consequence of General Relativity (GR) \cite{Penrose:1964wq,Hawking:1970zqf}. However, at very high energies, GR may be supplanted by a more fundamental theory that incorporates the quantum nature of spacetime. Since a complete theory of quantum gravity has not yet been established, another approach to addressing the singularity problem is to modify either the matter content or the gravitational sector at the classical level.

Several approaches have been proposed to construct nonsingular BHs without curvature singularities at their centers by introducing specific matter sources in 
four-dimensional spacetime.
In GR, nonlinear electrodynamics can give rise to regular 
BH solutions on a spherically symmetric and static background by choosing appropriate forms of the Lagrangian, ${\cal L}(F)$, where $F$ is the electromagnetic field strength \cite{Ayon-Beato:1998hmi,Ayon-Beato:2000mjt,Bronnikov:2000vy,Dymnikova:2004zc}.
However, it was recently shown that such nonsingular BHs are inevitably subject to Laplacian instabilities in the angular direction near a de Sitter center \cite{DeFelice:2024seu}. Even in more general theories 
where a scalar field $\phi$ with kinetic term $X$ is introduced in a Lagrangian of the form ${\cal L}(\phi,X,F)$, no linearly stable regular BH solutions have been found if the gravitational sector is described by the Einstein-Hilbert action \cite{DeFelice:2024ops}.

In cosmology, early attempts to resolve the initial Big Bang singularity include the pre-Big Bang and Ekpyrotic scenarios \cite{Veneziano:1991ek,Gasperini:1992em,Khoury:2001wf}, in which a period of cosmological contraction is followed by an expanding phase.
Both scenarios can be described within an effective 
four-dimensional framework of scalar-tensor theories, 
where the scalar degree of freedom corresponds to 
a dilaton field or the position of D-branes.
By taking into account higher-order corrections to the 
tree-level action, it is possible to realize regular bouncing solutions without curvature singularities \cite{Gasperini:1996fu,Brustein:1997cv,Foffa:1999dv,Cartier:1999vk,Tsujikawa:2002qc}. 
In the effective four-dimensional description, these nonsingular models fall within subclasses of Horndeski theories--the most general scalar-tensor theories with second-order field equations of motion \cite{Horndeski:1974wa,Deffayet:2011gz,Kobayashi:2011nu,Charmousis:2011bf}.

Numerous other examples of bouncing \cite{Qiu:2011cy,Easson:2011zy,Cai:2012va,Osipov:2013ssa,Qiu:2013eoa,Koehn:2013upa,Ijjas:2016tpn} 
or genesis \cite{Creminelli:2010ba,Creminelli:2012my,Hinterbichler:2012fr,Nishi:2015pta,Kobayashi:2015gga} 
cosmological solutions have been constructed within the framework of Horndeski theories. 
These solutions are typically plagued by the emergence 
of either singularities or ghost/Laplacian instabilities 
at early or late stages of cosmological evolution.
In Horndeski theories, there is a no-go theorem stating that, if a background solution is nonsingular during the entire cosmological evolution, the linear stability conditions for scalar and/or tensor perturbations are 
violated during some time interval \cite{Libanov:2016kfc,Kobayashi:2016xpl}. 
This no-go theorem can be circumvented by considering specific models in which unstable perturbations do not grow significantly within a short timescale---either by lowering the scale of ultraviolet completion~\cite{Pirtskhalava:2014esa} or by extending Horndeski theories to more general frameworks~\cite{Creminelli:2006xe, Pirtskhalava:2014esa, Koehn:2015vvy, deRham:2017aoj,Heisenberg:2018wye}. 
An alternative mechanism has been proposed to suppress the coefficients of second-order actions for scalar and/or tensor perturbations in the asymptotic 
past \cite{Ageeva:2020gti,Ageeva:2020buc,Ageeva:2021yik}. 
In the latter case, as long as the strong coupling energy scale is much higher than the scale associated with the classical background evolution, it is possible to construct linearly stable bouncing or genesis solutions.

Recently, geometric approaches have also been proposed to address BH and Big Bang singularities by incorporating an infinite tower of higher-curvature corrections \cite{Colleaux:2020wfv,Fernandes:2025fnz,Fernandes:2025eoc,Bueno:2024dgm,Bueno:2024eig}. 
If we restrict ourselves to theories that yield second-order field equations of motion to avoid Ostrogradsky instabilities, the Lagrangian should consist of a sum of the $n$-th order Lovelock curvature invariants 
${\cal R}^{(n)}$ \cite{Lovelock:1971yv,Lovelock:1972vz}, 
which preserve diffeomorphism invariance in 
$d$-dimensional spacetime.
For $2n \geq d$, the $n$-th order contributions 
to the field equations vanish identically. 
In four dimensions ($d=4$), the only nonvanishing contributions to the Lovelock Lagrangian are the cosmological constant term, ${\cal R}^{(0)}=1$, 
and the Ricci scalar, ${\cal R}^{(1)}=R$. 
The Gauss-Bonnet (GB) term, $R_{\rm GB}^2$,  corresponds to the second-order curvature 
invariant ${\cal R}^{(2)}=R_{\rm GB}^2$, 
which identically vanishes in four dimensions.
To extract nontrivial contributions from the GB Lagrangian
in four dimensions, one needs to either promote 
the coupling constant $\hat{\alpha}$ 
to a function of some dynamical degree of freedom 
(like a dilaton field) or rescale it in an appropriate manner.

By rescaling the coupling constant as 
$\hat{\alpha} \to \alpha/(d-4)$, one can extract nontrivial contributions of the higher-dimensional GB term in 
four dimensions.
This theory, originally proposed by Glavan and 
Lin \cite{Glavan:2019inb}, is commonly referred to 
as 4-dimensional Einstein-Gauss-Bonnet (4DEGB) gravity.
The original 4DEGB gravity suffers from several 
problems, including unphysical divergences in the perturbation equations \cite{Arrechea:2020evj,Arrechea:2020gjw}
and the lack of covariant field equations of a massless 
graviton \cite{Gurses:2020ofy,Gurses:2020rxb}. 

However, there are alternative approaches to deriving the four-dimensional effective action in 4DEGB gravity.
One such approach is conformal regularization, based on a rescaling of the metric tensor, 
$\tilde{g}_{\mu \nu}=e^{-2\phi}  g_{\mu \nu}$ \cite{Fernandes:2020nbq,Hennigar:2020lsl}, 
where $\phi$ is a scalar degree of freedom.
Another approach involves a Kaluza-Klein reduction of the higher-dimensional GB curvature invariant \cite{Lu:2020iav,Kobayashi:2020wqy,Ma:2020ufk}, 
in which the size of a maximally symmetric 
internal space is characterized by a scalar field. 
For a spatially flat internal space, the resulting 
4-dimensional effective action coincides with 
that obtained via conformal 
regularization\footnote{Note that this 4-dimensional effective action is also equivalent to the one arising from the trace anomaly due to closed loops of massless fields in an external gravitational background \cite{Riegert:1984kt}.}.
Indeed, the 4-dimensional action falls within a subclass of shift-symmetric Horndeski theories \cite{Lu:2020iav,Kobayashi:2020wqy,Ma:2020ufk}.

If we apply the scalar-tensor version of 4DEGB gravity 
to cosmology, the resulting solutions suffer from a strong coupling problem due to the vanishing kinetic term of the scalar field perturbation \cite{Kobayashi:2020wqy}. 
On a spherically symmetric and static background, 
an exact four-dimensional BH solution consistent with asymptotic flatness exists in 4DEGB gravity \cite{Glavan:2019inb}. 
However, the kinetic term of one of the even-parity perturbations vanishes everywhere, leading to the emergence of a strong coupling problem  \cite{Tsujikawa:2022lww}.
As discussed in Refs.~\cite{Bonifacio:2020vbk,Aoki:2020lig}, these strong coupling issues originate from the pathological behavior of scalar modes in the limit $d \to 4$.
This implies that the perturbative analysis becomes invalid due to the dominance of nonlinear perturbations. 

In Refs.~\cite{Colleaux:2020wfv,Fernandes:2025fnz}, 
a conformal regularization analogous to that in 4DEGB gravity was performed for arbitrary dimensions $d$ approaching the critical dimension $2n$. 
For each integer value of $n \geq 2$, 
a tower of Lagrangians ${\cal L}^{(n)}$ can be 
constructed by taking the limit $d \to 2n$.
A conformally regularized Lagrangian in four dimensions 
can then be obtained by summing over an infinite tower of
${\cal L}^{(n)}$ for $n=2,3,\cdots$. 
The resulting four-dimensional Lagrangian belongs to a subclass of shift-symmetric Horndeski theories. 
When applied to a spatially flat Friedmann-Lema\^{i}tre-Robertson-Walker (FLRW) background with matter fields, this theory admits an inflationary solution characterized by a nearly constant Hubble expansion rate $H$, which replaces the Big Bang singularity \cite{Fernandes:2025fnz}.
In $(2+1)$-dimensional spacetime, a similar regularization procedure--based on summing an infinite tower of curvature corrections--yields nonsingular BH solutions without curvature singularities \cite{Fernandes:2025eoc}.

In this paper, we investigate the behavior of perturbations 
on a spatially flat FLRW background in four-dimensional theories derived from an infinite sum of the 
regularized Lagrangian ${\cal L}^{(n)}$. 
The primary objective is to elucidate whether strong coupling or ghost/Laplacian instabilities arise when the infinite tower of corrections is taken into account.
Indeed, we show that the background solution 
$\dot{\phi}=H$, which is responsible for inflation, 
suffers from a severe strong coupling problem 
of the vanishing of the scalar field 
perturbation throughout the cosmological evolution. 
Moreover, tensor perturbations are subject to 
Laplacian instabilities during inflation.
Even when $\dot{\phi}$ differs from $H$, Laplacian instabilities of tensor perturbations still persist, 
along with the divergence of the kinetic term for scalar perturbations and the scalar sound speed 
at the onset of inflation.
Thus, the background inflationary solution fails to provide a physically consistent description due to the dominance of inhomogeneities.

This paper is organized as follows. 
In Sec.~\ref{4Dsec}, we briefly review the regularized four-dimensional Lovelock gravity constructed from an infinite sum of curvature corrections.
In Sec.~\ref{inssec}, we discuss the background dynamics of inflation that replaces the Big Bang singularity by 
selecting various sets of coefficients $c_n$ for each regularized Lagrangian. 
In Sec.~\ref{stosec}, we analyze the behavior of perturbations in regularized four-dimensional Lovelock gravity and show how the strong coupling problem in the scalar sector and instabilities in the tensor sector 
emerge in this scenario.
Sec.~\ref{consec} is devoted to conclusions. 
In Appendix, we present the behavior of the background and perturbations for the choice of coefficients 
$c_n=1/n$.

%%%%%%%%%%%%%%%%%%%%%%%%%%%%%%%%%%%%%
\section{Regularized four-dimensional 
Lovelock gravity}
\label{4Dsec}
%%%%%%%%%%%%%%%%%%%%%%%%%%%%%%%%%%%%%

The Lovelock gravitational action, which is defined in 
$d$-spacetime dimensions and is invariant under 
$d$-dimensional diffeomorphisms, is constructed 
to yield second-order field equations of motion. 
The action is expressed in 
the form ${\cal S}_{\rm L}^{(n)}=\int {\rm d}^d x \sqrt{-g}\,
\alpha_n {\cal R}^{(n)}$, where $g$ is the determinant 
of the metric tensor $g_{\mu \nu}$, 
$\alpha_n$ is a constant, and ${\cal R}^{(n)}$ is 
the $n$-th order curvature invariant defined by 
\be
{\cal R}^{(n)} \equiv \frac{1}{2^n} 
\delta^{\mu_1 \nu_1 \cdots \mu_n \nu_n}_{\alpha_1 
\beta_1 \cdots \alpha_n \beta_n} 
\prod_{i=1}^{n} {R^{\alpha_i \beta_i}}_{\mu_i \nu_i}\,,
\label{calRn}
\ee
where 
\be
\delta^{\mu_1 \nu_1 \cdots \mu_n \nu_n}_{\alpha_1 
\beta_1 \cdots \alpha_n \beta_n} 
\equiv n! \delta^{\mu_1}_{[\alpha_1} 
\delta^{\nu_1}_{\beta_1} \cdots 
\delta^{\mu_n}_{\alpha_n} \delta^{\nu_n}_{\beta_n]}\,,
\ee
is the generalized Kronecker delta, and 
${R^{\alpha_i \beta_i}}_{\mu_i \nu_i}$ is the Riemann 
tensor. 

The dimensional regularization method adopted in 
Refs.~\cite{Fernandes:2020nbq,Hennigar:2020lsl,Colleaux:2020wfv,Fernandes:2025fnz} is based on a conformal rescaling of the metric, 
$\tilde{g}_{\mu \nu}=e^{-2\phi} g_{\mu \nu}$, where 
$\phi$ is a scalar degree of freedom. 
Moreover, the following limit is taken at each order 
of the Lovelock scalar:
\be
{\cal L}^{(n)}=\lim_{d \to 2n} 
\frac{\sqrt{-g}{\cal R}^{(n)}-\sqrt{-\tilde{g}} 
\tilde{\cal R}^{(n)}}{d-2n}\,, 
\label{Ln}
\ee
where a tilde represents quantities in the frame 
with the metric tensor $\tilde{g}_{\mu \nu}$. 
The 4DEGB gravity corresponds to $n=2$, 
in which case the contribution of the GB term 
${\cal R}^{(2)}=R_{\rm GB}^2$ is extracted 
via a conformal rescaling with a factorization involving division by $d-4$. In Ref.~\cite{Colleaux:2020wfv}, 
the above conformal regularization is applied to 
other integer values of $n$ larger than 2. 
For $n \geq 3$, the limit $d \to 2n$ corresponds to 
the spacetime dimension greater than $d=6$. 
Refs.~\cite{Colleaux:2020wfv,Fernandes:2025fnz} 
incorporate the Lagrangians ${\cal L}^{(n)}$ for $n \geq 3$ 
into four-dimensional theories, in addition to the 
4DEGB contribution.
Taking into account the Ricci scalar $R$ and
the cosmological constant $\Lambda$, the  
four-dimensional theory obtained from the infinite 
sum of ${\cal L}^{(n)}$ over $n=2,3,\cdots$ 
is given by the action 
\ba
{\cal S} &=& 
\int {\rm d}^4x\sqrt{-g}  
\left[
\frac{1}{2} R -\Lambda + \frac{1}{2\ell^2} 
\sum_{n=2}^{\infty} c_n \ell^{2n} 
{\cal L}^{(n)}\right] \nonumber \\
&&+{\cal S}_m (g_{\mu \nu}, \Psi_m)\,,
\label{action}
\ea
where $\ell$ is a constant with dimensions of length, 
$c_n$ are coefficients dependent on $n$, and 
\ba
{\cal L}^{(n)} &=& 
G_2^{(n)}(X)-G_{3}^{(n)}(X)\square\phi 
+G_{4}^{(n)}(X) R \nonumber \\
& &+G_{4,X}^{(n)}(X) \left[ (\square \phi)^{2}
-(\nabla_{\mu}\nabla_{\nu} \phi)
(\nabla^{\mu}\nabla^{\nu} \phi) \right] \nonumber \\
& &
+G_{5}^{(n)}(X) G_{\mu \nu} \nabla^{\mu}\nabla^{\nu} \phi
\nonumber \\
& &
-\frac{1}{6}G_{5,X}^{(n)}(X)
[ (\square \phi )^{3}-3(\square \phi)\,
(\nabla_{\mu}\nabla_{\nu} \phi)
(\nabla^{\mu}\nabla^{\nu} \phi) \nonumber \\
& &
+2(\nabla^{\mu}\nabla_{\alpha} \phi)
(\nabla^{\alpha}\nabla_{\beta} \phi)
(\nabla^{\beta}\nabla_{\mu} \phi) ]  \,,
\label{Ln}
\ea
with $G_{\mu \nu}$ being the four-dimensional 
Einstein tensor, and \cite{Fernandes:2025fnz}
\ba
G_2^{(n)}(X) &=& 2^{n+1} (n-1)(2n-3)X^n\,,
\label{Gchoice0} \nonumber \\
G_3^{(n)}(X) &=& -2^n n (2n-3) X^{n-1}\,,\nonumber \\
G_4^{(n)}(X) &=& 2^{n-1} n X^{n-1}\,,\nonumber \\
G_5^{(n)}(X) &=& 
\begin{cases}
-4 \ln X& \quad ({\rm for}~n=2)\,, \nonumber \\
-2^{n-1} \dfrac{n(n-1)}{n-2} 
X^{n-2} & \quad ({\rm for}~n \geq 3)\,. 
\end{cases} \nonumber \\
\label{Gchoice}
\ea
The coupling functions $G_i^{(n)}$ (with $i=2,3,4,5$) 
depend only on the scalar kinetic term 
$X=-(1/2)g^{\mu \nu}\nabla_{\mu}\phi \nabla_{\nu}\phi$, 
where $\nabla_{\mu}$ denotes the covariant derivative. 
Since such theories remain invariant under a constant shift of the scalar field, $\phi \to \phi+c$, they belong to 
the subclass of shift-symmetic Horndeski theories.
We adopt the notations $G_{i,X}^{(n)}(X)
=\partial G_i^{(n)}/\partial X$ and 
$\square \phi= g^{\mu \nu} \nabla_{\mu} \nabla_{\nu} 
\phi$, and work in units where the reduced Planck mass
$M_{\rm pl}$ is set to 1. 
We will consider the dynamics on a time-dependent, isotropic cosmological background, in which case 
$X$ is positive\footnote{If we consider other backgrounds where $X$ is negative, such as a spherically symmetric 
and static background, we can generalize the function 
$G_{5}^{(2)}(X)=-4 \ln X$ to  $G_{5}^{(2)}(X)=-4 \ln |X|$.}.
The action ${\cal S}_m$ describes the matter fields 
$\Psi_m$, which are assumed 
to be minimally coupled to gravity.

Taking the infinite sum of $G_i^{(n)}$ over 
$n=2,3,\cdots$, the resulting theory is 
characterized by the following coupling functions:
\ba
G_{2}(X) &= & 
-\Lambda+\frac{1}{2 \ell^2} 
\sum_{n=2}^{\infty} 
c_n \ell^{2n} G_2^{(n)}(X)\,,\label{G2} \\
G_{3}(X) &=& 
\frac{1}{2\ell^2}\sum_{n=2}^{\infty} 
c_n \ell^{2n} G_3^{(n)}(X)\,,\\
G_{4}(X) &=& \frac{1}{2} + \frac{1}{2\ell^2}\sum_{n=2}^{\infty} 
c_n \ell^{2n} G_4^{(n)}(X) \,,\\
G_{5}(X) &=&
\frac{1}{2\ell^2}\sum_{n=2}^{\infty} 
c_n \ell^{2n} G_5^{(n)}(X)\,.
\label{G5}
\ea
The total action is given by 
${\cal S}=\int {\rm d}^4 x \sqrt{-g}\,{\cal L}+{\cal S}_m$, 
where ${\cal L}$ is obtained by replacing $G_{i}^{(n)}(X)$  
with $G_{i}(X)$ (for $i=2,3,4,5$) in Eq.~(\ref{Ln}).

The coupling functions in Eqs.~(\ref{G2})-(\ref{G5}) 
depend on the coefficients $c_n$. 
If we choose
\be
c_n=1 \quad {\rm for~all~}n 
\qquad ({\rm Model~1})\,,
\ee
it follows that
\ba
\hspace{-0.9cm}
G_2(X) &=& -\Lambda+\frac{4 \ell^2 X^2 
(1+6 \ell^2 X)}{(1-2\ell^2 X)^3}, \nonumber \\
\hspace{-0.9cm}
G_3(X) &=& \frac{1-10 \ell^2 X}{(1-2\ell^2 X)^3}\,, \nonumber \\
\hspace{-0.9cm}
G_4(X) &=& \frac{1}{2(1-2\ell^2 X)^2}\,,\nonumber \\
\hspace{-0.9cm}
G_5(X) &=& -2 \ell^2 \left[ \frac{1-2\ell^4 X^2}
{(1-2\ell^2 X)^2}+\ln \left( \frac{2\ell^2 X}{1-2\ell^2 X} 
\right) \right],
\label{example1}
\ea
which coincide with those obtained 
in Ref.~\cite{Fernandes:2025fnz}. 
For the choice
\be
c_n=\frac{1-(-1)^n}{2n} \qquad ({\rm Model~2})\,,
\ee
we have\footnote{
The cubic Horndeski function can also be written 
in the form 
$G_3=(1-20\ell^4 X^2)/(1-4\ell^4 X^2)^2$ \cite{Fernandes:2025fnz}.
This expression differs from Eq.~(\ref{example2}) only  
by a constant factor, and thus both forms of 
$G_3$ yield the same field equations of motion.}
\ba
G_2(X) &=& -\Lambda+\frac{2 X(28 \ell^4 X^2-3)}
{(1-4\ell^4 X^2)^2}+\frac{3}{\ell^2} 
\tanh^{-1} (2\ell^2 X), \nonumber \\
G_3(X) &=& -\frac{4 \ell^4 X^2 (3+4\ell^4 X^2)}
{(1-4\ell^4 X^2)^2}\,, \nonumber \\
G_4(X) &=& \frac{1}{2(1-4\ell^4 X^2)}\,,\nonumber \\
G_5(X) &=& -\frac{2 \ell^4 X}{1-4\ell^4 X^2}
-\ell^2 \tanh^{-1} (2\ell^2 X)\,.
\label{example2}
\ea
In Appendix, we also present the Horndeski functions for the alternative choice of coefficients $c_n=1/n$, which we refer to as Model 3.

%%%%%%%%%%%%%%%%%%%%%%%%%%%%%
\section{Inflationary solutions replacing 
the Big Bang singularity} 
\label{inssec}
%%%%%%%%%%%%%%%%%%%%%%%%%%%%%

To study the background cosmological dynamics,  
we consider a spatially flat FLRW spacetime 
described by the four-dimensional line element 
\be
{\rm d}s^2=-{\rm d}t^2+a^2(t) \delta_{ij} 
{\rm d}x^i {\rm d}x^j\,,
\label{FLRW}
\ee
where $a(t)$ is a time-dependent scale factor. 
The choice of this line element is justified when the 
inhomogeneity and anisotropy of the Universe are sufficiently small.
We assume that the action ${\cal S}_m$ in the matter sector 
corresponds to that of a perfect fluid with background energy density $\rho_m$ and pressure $P_m$.
The Hubble expansion rate is defined as $H=\dot{a}/a$, 
where a dot denotes a derivative with respect to 
cosmic time $t$. The background equations of motion 
are given by \cite{Kobayashi:2011nu,DeFelice:2011hq,DeFelice:2011bh,Kase:2018aps}
\ba
& &
6G_4 H^2+G_2-\dot{\phi}^2 G_{2,X}
-3H \dot{\phi}^3 G_{3,X}  
\nonumber \\
& &
-6H^2 \dot{\phi}^2 (2G_{4,X}+\dot{\phi}^2 
G_{4,XX})  \nonumber \\
& &
-H^3 \dot{\phi}^3 (5G_{5,X}+\dot{\phi}^2 
G_{5,XX})=\rho_m\,,  \label{back1} \\
& &
2q_t \dot{H}-D_6 \ddot{\phi}+D_7 \dot{\phi}
=-\rho_m-P_m\,, \label{back2} \\
&& 
\frac{\rd }{\rd t} \left( a^3 J \right)=0\,, 
\label{back3} \\
& &
\dot{\rho}_m+3H \left( \rho_m+P_m \right)=0\,,
\label{back4} 
\ea
where\footnote{We follow the notations of 
Ref.~\cite{Kase:2018aps}, except that the sign 
of $G_3$ is opposite to that used in this paper. 
For the expression of the scalar field current $J$, 
readers may refer to Eq.~(3.12) of Ref.~\cite{Kobayashi:2011nu}.}
\ba
q_t &=& 2G_4-2\dot{\phi}^2 G_{4,X}
-H \dot{\phi}^3 G_{5,X}\,, 
\label{qt} \\
D_6 &=& \dot{\phi}^2 G_{3,X}
+4H \dot{\phi} (G_{4,X}+\dot{\phi}^2 G_{4,XX})  \nonumber \\
& &
+H^2 \dot{\phi}^2 
( 3G_{5,X}+\dot{\phi}^2 G_{5,XX} )\,,\\
D_7 &=& \dot{\phi} G_{2,X}+3H \dot{\phi}^2 G_{3,X}
+6H^2 \dot{\phi} (G_{4,X}+\dot{\phi}^2 G_{4,XX}) 
\nonumber \\
& &
+H^3 \dot{\phi}^2 
( 3G_{5,X}+\dot{\phi}^2 G_{5,XX} )\,,\\
J &= &\dot{\phi} G_{2,X}+3H \dot{\phi}^2 G_{3,X}
+6 H^2 \dot{\phi} (G_{4,X}+\dot{\phi}^2 
G_{4,XX}) \nonumber \\
& &
+H^3 \dot{\phi}^2 (3G_{5,X}+\dot{\phi}^2 G_{5,XX})\,.
\label{Jexpression}
\ea
The cosmological constant $\Lambda$ appears in 
Eq.~(\ref{back1}) through the coupling $G_2$.
Note that $J$ corresponds to the scalar field current. 
The right-hand side of Eq.~(\ref{back3}) vanishes, reflecting the fact that the four-dimensional action (\ref{action}) is invariant under the shift $\phi \to \phi+c$. 
In shift-symmetric Horndeski theories, the coefficient 
$D_7$ coincides with $J$. 
However, this property does not hold in the most general Horndeski theories, where the coupling 
functions $G_i$ depend on both $\phi$ and $X$.
Equation (\ref{back3}) can be integrated to give
\be
J=\frac{{\cal C}}{a^3}\,,
\label{Jre}
\ee
where ${\cal C}$ is an integration constant. 

Among Eqs.~(\ref{back1})-(\ref{back4}), three of them 
are independent. Indeed, taking the time derivative of 
(\ref{back1}) and using (\ref{back2}) and 
(\ref{back4}) to eliminate $\dot{\rho}_m$, 
$\rho_m$, and $P_m$, 
the resulting equation is consistent with (\ref{back3}).
Solving Eqs.~(\ref{back2}) and (\ref{back3}) for 
$\dot{H}$ and $\ddot{\phi}$, it follows that 
\ba
\hspace{-0.9cm}
\dot{H} &=& -\frac{1}{q_s} \left[ D_7 (2D_1 \dot{\phi}
+3D_6 H)+2D_1 (\rho_m+P_m) \right],
\label{dotH} \\
\hspace{-0.9cm}
\ddot{\phi} &=& \frac{1}{q_s} \left[ 
3D_7 (D_6 \dot{\phi}-2H q_t)+3D_6 
(\rho_m+P_m) \right],
\label{ddotphi}
\ea
where 
\be
q_s=4D_1 q_t+3D_6^2\,,
\label{qsdef}
\ee
with 
\ba
\hspace{-0.4cm}
D_1 &=& 
\frac{1}{2} (G_{2,X}+\dot{\phi}^2 G_{2,XX})
+\frac{3}{2}H \dot{\phi} (2G_{3,X}+\dot{\phi}^2 G_{3,XX}) 
\nonumber \\
\hspace{-0.4cm}
& &+3H^2 (G_{4,X}+4\dot{\phi}^2 G_{4,XX}
+\dot{\phi}^4 G_{4,XXX})\nonumber \\
\hspace{-0.4cm}
& & 
+\frac{1}{2} H^3 \dot{\phi} (6 G_{5,X}+7 \dot{\phi}^2 
G_{5,XX}+\dot{\phi}^4 G_{5,XXX})\,.
\ea
For $q_s>0$, as long as the numerators of Eqs.~(\ref{dotH}) 
and (\ref{ddotphi}) remain finite, it is possible to avoid singularities in the background equations of motion.

In what follows, we discuss the background cosmological dynamics for models with two different choices of $c_n$. 
Since we are primarily interested in the dynamics 
of the early Universe, we include radiation, described 
by the equation of state parameter
\be
w_m=\frac{P_m}{\rho_m}=\frac{1}{3}\,.
\ee
In this case, the radiation energy density evolves 
according to
\be
\rho_m=\rho_{mi} a^{-4}\,,
\label{rhomso}
\ee
where $\rho_{mi}$ is a constant.
The cosmological constant $\Lambda$ is relevant only to the late-time dynamics associated with dark energy, 
so we set 
\be
\Lambda=0\,,
\ee
in the following discussion. 

\subsection{Model 1}
\label{model1sec}

In Model 1, the Horndeski functions are given by Eq.~(\ref{example1}). 
In this case, the scalar field equation 
takes the form 
\be
J=\frac{4 \ell^2 (1+5\ell^2 \dot{\phi}^2)}
{(1-\ell^2 \dot{\phi}^2)^4} (\dot{\phi}-H)^3
=\frac{{\cal C}}{a^3}\,.
\label{Jmo1}
\ee
As long as 
\be
{\cal C}=0\,,
\ee
Eq.~(\ref{Jmo1}) admits the following solution 
\be
\dot{\phi}=H\,.
\label{atra}
\ee
On this background solution, Eqs.~(\ref{back1}), (\ref{dotH}), 
and (\ref{ddotphi}) simplify to
\ba
& &
\frac{3 H^2}{1-\ell^2 H^2}=\rho_m\,,
\label{atback3}\\
& &
\dot{H}=\ddot{\phi}=-\frac{2}{3}(1-\ell^2 H^2)^2
\rho_m\,.
\label{atback4}
\ea
From Eq.~(\ref{atback3}), a consistent solution 
with $\rho_m>0$ exists only in the regime $\ell H<1$. 
Using Eqs.~(\ref{rhomso}) and (\ref{atback3}) 
in an expanding Universe ($H>0$), we find
\be
H=\frac{1}{\ell} \sqrt{\frac{a_{m_1}^4}
{a^4+a_{m_1}^4}}\,,
\label{Hsomodel1}
\ee
where $a_{m_1}=(\ell^2 \rho_{mi}/3)^{1/4}$. 
For $a \ll a_{m_1}$, $H$ is nearly constant and
close to $\ell^{-1}$.  
This corresponds to the inflationary stage, characterized by the evolution of the scale factor as $a \propto \exp (Ht)$. 
Inflation is driven by an almost constant 
scalar field derivative $\dot{\phi}$.
From Eq.~(\ref{atback4}), we also find that 
$\dot{H}$ is close to 0 around $H \simeq \ell^{-1}$.

The inflationary period ends when $a$ grows 
larger than $a_{m_1}$, after which   
$H \simeq \sqrt{\rho_{mi}}/(\sqrt{3} a^2)$ 
and $\dot{H} \simeq -2\rho_{mi}/(3 a^4)$. 
This corresponds to the radiation-dominated era, characterized by the scale factor evolving as 
$a \propto t^{1/2}$. 
Since $H$ is bounded from above ($H<\ell^{-1}$), 
the conventional Big Bang singularity, in which 
$H$ diverges as $a \to 0$, is absent in this scenario. 
Instead, the Universe starts from the 
inflationary stage followed by the radiation era. 
Taking the limit $a \to 0$, the right-hand side of 
Eq.~(\ref{atback3}) diverges as $\rho_m \propto a^{-4}$. 
In the same limit, $H$ approaches $\ell^{-1}$ 
according to the relation 
$1-\ell^2 H^2 \propto a^{4}$. 
Thus, the finiteness of $H$ is realized by the presence 
of a matter fluid whose energy density is unbounded 
from above.

The above discussion is based on the choice ${\cal C}=0$ 
in Eq.~(\ref{Jre}). If ${\cal C} \neq 0$, as long as 
$\dot{\phi}$ remains close to $H$, it should be possible to realize an inflationary solution similar to the one 
discussed above. To accommodate such a case, 
we write $\dot{\phi}$ in the form 
\be
\dot{\phi}=H \left[ 1+\epsilon(t) \right]\,,
\label{phiHre}
\ee
where $\epsilon(t)$ is a time-dependent function satisfying 
the condition $|\epsilon(t)| \ll 1$. 
We substitute Eq.~(\ref{phiHre}) into the left-hand side of 
Eq.~(\ref{Jmo1}) and perform an expansion with 
respect to the small parameter $\epsilon(t)$. 
By keeping only the leading-order term, it follows that
\be
|\epsilon(t)| \simeq
\left( \frac{|{\cal C}|}
{4\ell^2 (1+5\ell^2 H^2)} \right)^{1/3}
\frac{(1-\ell^2 H^2)^{4/3}}{a}  \frac{1}{H}\,.
\label{ept}
\ee
At leading order in the expansion in $\epsilon(t)$, 
we obtain the same background equations as 
Eqs.~(\ref{atback3}) and (\ref{atback4}). 
Substituting the leading-order 
solution (\ref{Hsomodel1})
into the term $(1-\ell^2 H^2)^{4/3}$ in Eq.~(\ref{ept}), 
we obtain  
\be
|\epsilon(t)| \simeq
\left( \frac{|{\cal C}|}
{4\ell^2 (1+5\ell^2 H^2)} \right)^{1/3}
\frac{a^{13/3}}{(a^4+a_{m_1}^4)^{4/3}}
 \frac{1}{H}\,.
\label{diffe}
\ee
During inflation ($a \ll a_{m_1}$), 
we have $|\epsilon(t)| \propto a^{13/3} \propto e^{13Ht/3}$,  
and hence the difference between $\dot{\phi}$ and 
$H$ increases exponentially. 
For the validity of using Eq.~(\ref{Hsomodel1}) as the leading-order solution to $H$, we need to choose 
the value of ${\cal C}$ such that $|\epsilon (t)| \ll 1$ 
during inflation. 
Since the order of $|\epsilon(t)|$ at the end of inflation 
(time $t_f$) can be estimated as
$|\epsilon(t_f)| \approx (|{\cal C}|/\ell^2)^{1/3} \ell/a_{m_1}$, 
we require that $|{\cal C}| \ll a_{m_1}^3/\ell$ to ensure 
the condition $|\epsilon (t)| \ll 1$ during inflation.

After the Universe enters the radiation-dominated epoch 
($a \gtrsim a_{m_1}$ and $H \ll \ell^{-1}$), 
$|\epsilon(t)|$ is proportional to
$(aH)^{-1} \propto t^{1/2}$. 
From Eq.~(\ref{Jmo1}), the same parameter also increases 
as $|\epsilon(t)| \propto (aH)^{-1} \propto t^{1/3}$ 
during the subsequent matter-dominated era.
This indicates that $|{\cal C}|$ must be extremely small 
to satisfy the condition $|\epsilon(t)| \ll 1$ throughout the cosmological evolution.
If $\dot{\phi}$ begins to deviate from $H$ during the decelerating cosmological epochs, it happens that 
$\dot{\phi}$ evolves more slowly than $H \propto t^{-1}$. 
In this case, the standard radiation and matter eras can 
be disrupted by the dominance of the scalar field energy density.
As long as $|\epsilon(t)| \ll 1$, the background dynamics 
during the radiation era is approximated by
Eqs.~(\ref{atback3}) and (\ref{atback4}), 
with $H \ll \ell^{-1}$.

\subsection{Model 2}
\label{model2sec}

In Model 2, defined by the Horndeski functions given 
in Eq.~(\ref{example2}), the scalar field equation 
can be written in the form
\be
J=\frac{4 \ell^4  \dot{\phi}^2 (3+5 \ell^4 \dot{\phi}^4)}
{(1-\ell^4 \dot{\phi}^4)^3} (\dot{\phi}-H)^3
=\frac{{\cal C}}{a^3}\,.
\label{Jex}
\ee
For the choice ${\cal C}=0$, we obtain 
the same solution as in Eq.~(\ref{atra}). 
There also exists another branch with $\dot{\phi}=0$, 
but in this case, the scalar field does not contribute to the cosmological dynamics. 
For the solution $\dot{\phi}=H$, the background 
equations simplify to
\ba 
& &
\frac{3}{2\ell^2} \ln \left( 
\frac{1+\ell^2 H^2}{1-\ell^2 H^2}
\right)=\rho_m\,,
\label{atback1} \\
& &
\dot{H}=\ddot{\phi}=-\frac{2}{3}(1-\ell^4 H^4)
\rho_m\,.
\label{atback2} 
\ea
From Eq.~(\ref{atback1}), a consistent background 
solution exists only when $\ell H<1$.
By solving Eq.~(\ref{atback1}) for $H~(>0)$, 
using Eq.~(\ref{rhomso}), and defining 
$a_{m_2}=(2\ell^2 \rho_{mi}/3)^{1/4}$, we obtain
\be
H=\frac{1}{\ell} \sqrt{\frac{\exp( a_{m_2}^4/a^4)-1}
{\exp( a_{m_2}^4/a^4)+1}}\,,
\label{Heq}
\ee
so that $H$ is bounded from above ($H<\ell^{-1}$).

When the scale factor is in the regime $a \ll a_{m_2}$, 
inflation occurs with a nearly constant Hubble 
expansion rate $H$ close to $\ell^{-1}$. 
After $a$ grows larger than $a_{m_2}$, 
expanding the term $\exp(a_{m_2}^4/a^4)$ 
in Eq.~(\ref{Heq}) yields the approximate relation 
$H \simeq \sqrt{\rho_{mi}}/(\sqrt{3}a^2)$. 
Thus, the inflationary period is followed by the radiation 
era, characterized by the scale factor evolving 
as $a \propto t^{1/2}$. 
As in Model 1, $H$ approaches the finite constant $\ell^{-1}$ 
toward the asymptotic past, so the 
curvature singularity is absent.

For ${\cal C} \neq 0$, we express the difference between 
$\dot{\phi}$ and $H$ in the form given by Eq.~(\ref{phiHre}). 
Performing an expansion of Eq.~(\ref{Jex}) 
in the small parameter 
$\epsilon(t)$ and using Eq.~(\ref{Heq}) as the 
leading-order solution for $H$, it follows that 
\be
|\epsilon (t)| \simeq \left( \frac{16 |{\cal C}|}
{\ell^4 (3+5 \ell^4 H^4)}  \right)^{1/3}
\frac{\exp( a_{m_2}^4/a^4)}
{[\exp( a_{m_2}^4/a^4)+1]^2} \frac{1}{aH^{5/3}}\,.
\label{Hphi2}
\ee
In the limit $a \to 0$, $|\epsilon (t)|$ rapidly 
approaches 0. 
During inflation ($a \lesssim a_{m_2}$), $|\epsilon (t)|$ 
increases proportionally to 
$a^{-1} \exp(-a_{m_2}^4/a^4)$.
To ensure the condition $|\epsilon (t)| \ll 1$ by the end 
of inflation, we require that $|{\cal C}| \ll a_{m_2}^3/\ell$. 
During the subsequent radiation era, $|\epsilon (t)|$ 
grows as $|\epsilon (t)| \propto a^{-1}H^{-5/3} \propto t^{7/6}$. 
Thus, we need to choose ${\cal C}$ to be very close to 0 
to maintain the condition $|\epsilon (t)| \ll 1$
throughout the cosmological evolution.
The continuous increase in $|\epsilon(t)|$ is similar to that observed in Model 1, although the growth rates 
are different.

\subsection{Models of each power $n$}
\label{modeln}

In both Models 1 and 2, we have shown that  
the solution $\dot{\phi}=H$ drives cosmic inflation 
in the regime where $H$ is close to $\ell^{-1}$. 
As we will see in Appendix, the same property also 
holds for the choice $c_n=1/n$. 
To understand why the solution $\dot{\phi}=H$ 
always exists, we consider the coupling functions
$G_{2,3,4,5}^{(n)}$ for each $n$ as given in 
Eqs.~(\ref{G2})-(\ref{G5}), 
e.g., $G_3(X)=(2\ell^2)^{-1}\,c_n \ell^{2n} G_{3}^{(n)}(X)$. 
Then, the scalar field equation takes the form  
\be
J=2 c_n n (n-1)(2n-3) \ell^2
(\ell \dot{\phi})^{2(n-2)} (\dot{\phi}-H)^3
=\frac{{\cal C}}{a^3}\,.
\label{Jeach}
\ee
This implies that the solution $\dot{\phi}=H$ 
exists when ${\cal C}=0$.
It is evident that this solution continues to exist 
even when the infinite sum over $n=2,3,\cdots$ 
is taken. At each order in $n$, the background equations 
of motion evaluated on the solution $\dot{\phi}=H$ 
are given by 
\ba
& &
3H^2 \left[ 1+c_n (\ell H)^{2n-2} \right]=\rho_m\,,
\label{backn1}\\
& &
\dot{H}=\ddot{\phi}=-\frac{2\rho_m}
{3 [1+c_n n (\ell H)^{2n-2}]}\,.
\label{backn2}
\ea
In the regime where $|c_n| (\ell H)^{2n-2} \gg 1$, we require 
that $c_n>0$ for the consistency of Eq.~(\ref{backn1}).
In this case, as $a \to 0$, we observe that $H$ grows 
without bound toward the asymptotic past, 
along with the divergence of $\rho_m \propto a^{-4}$. 
Thus, the curvature singularity is generally present 
in theories with finite $n$. 
In the low-energy regime characterized by 
$c_n n (\ell H)^{2n-2} \ll 1$, the standard 
radiation era with $\dot{H} \simeq -2\rho_m/3$
is recovered.

The above results indicate that summing over the infinite series of coupling functions $G_{2,3,4,5}^{(n)}$ 
over $n$ is essential for achieving inflation with 
a finite Hubble expansion rate $H<\ell^{-1}$. 
As we have seen in Secs.~\ref{model1sec} 
and \ref{model2sec}, even though $\rho_m$ diverges 
as $a \to 0$, the Hubble expansion rate remains finite 
in Models 1 and 2. This property also holds in Model 3, 
see Appendix for details.
 
%%%%%%%%%%%%%%%%%%%%%%%%%%%%%
\section{Linear cosmological perturbations} 
\label{stosec}
%%%%%%%%%%%%%%%%%%%%%%%%%%%%%

In this section, we study the behavior of linear 
cosmological perturbations 
on the spatially flat FLRW background (\ref{FLRW}). 
We consider the four-dimensional perturbed line element given by 
\ba
\hspace{-0.5cm}
{\rm d}s^2 
&=& -(1+2\alpha) {\rm d}t^2+2 \partial_i \chi 
{\rm d} t {\rm d}x^i \nonumber \\
\hspace{-0.5cm}
& &
+a^2(t) \left[ (1+2\zeta) \delta_{ij}
+2 \partial_i \partial_j E+h_{ij} \right]{\rm d}x^i {\rm d}x^j\,,
\label{permet}
\ea
where $\alpha$, $\chi$, $\zeta$, and $E$ 
are scalar metric perturbations, and $h_{ij}$ is 
the tensor perturbation satisfying the traceless and 
transverse conditions ${h^{i}}_i=0$ and $\partial_i h^{ij}=0$. 
The perturbed fields depend on time $t$ and 
spatial coordinates $x^i$, where we use the notation 
$\partial_i=\partial/\partial x^i$. 
Since vector perturbations are nondynamical in 
scalar-tensor theories, we omit them from our analysis.
The scalar field $\phi$ and the matter density $\rho_m$ 
are decomposed into the background and perturbed 
parts, as $\phi=\bar{\phi}(t)+\delta \phi(t,x^i)$ and 
$\rho_m=\bar{\rho}_m(t)+\delta \rho_m(t,x^i)$, where 
we will drop the overbar for brevity in what follows.

For tensor perturbations propagating along the 
$z$ direction, we can choose the components of $h_{ij}$ 
as $h_{11}=-h_{22}=h_1(t,z)$ and 
$h_{12}=h_{21}=h_2(t,z)$. 
In Fourier space with comoving wavenumber $k$, the 
second-order action for tensor modes can be 
written as \cite{Kobayashi:2011nu,Kase:2018aps}
\be
{\cal S}_t^{(2)}=\int \frac{{\rm d}t  {\rm d}^3k}
{(2 \pi)^3} \sum_{i=1}^{2} \frac{a^3}{4}q_t
\left( \dot{h}_i^2-c_t^2 \frac{k^2}{a^2} h_i^2 
\right)\,,
\ee
where $q_t$ is defined by Eq.~(\ref{qt}), and $c_t^2$ is 
the squared tensor propagation speed given by 
\be
c_t^2=\frac{2G_4-\dot{\phi}^2 \ddot{\phi}\,G_{5,X}}
{2G_4-2\dot{\phi}^2 G_{4,X}
-H \dot{\phi}^3 G_{5,X}}\,.
\label{ct2}
\ee
To avoid ghost and Laplacian instabilities in tensor perturbations, the following conditions 
must be satisfied:
\be
q_t>0\,,\qquad c_t^2>0\,.
\ee

In the presence of a perfect fluid, the second-order action 
for scalar perturbations and the corresponding linear perturbation equations of motion are presented 
in a gauge-ready form in Ref.~\cite{Kase:2018aps} 
(see also Refs.~\cite{DeFelice:2011hq,DeFelice:2011bh}).
We adopt the unitary gauge, which is characterized by 
\be
\delta \phi=0\,,\qquad E=0\,.
\ee
We then eliminate the nondynamical perturbations 
$\alpha$, $\chi$, and the fluid velocity potential $v$ 
from the quadratic-order action by using their equations of motion. In Fourier space, the second-order action for the two dynamical perturbations $\zeta$ and $\delta \rho_m$ 
can be written in the form 
\be
{\cal S}_{s}^{(2)}=\int \frac{{\rm d}t  {\rm d}^3k}
{(2 \pi)^3}\,a^{3}
\left( 
\dot{\vec{\mathcal{X}}}^{t}{\bm K}\dot{\vec{\mathcal{X}}}
-\vec{\mathcal{X}}^{t} \tilde{{\bm G}}\vec{\mathcal{X}}
-\vec{\mathcal{X}}^{t}{\bm B}\dot{\vec{\mathcal{X}}}
\right)\,,
\label{Ss2}
\ee
where ${\bm K}$, $\tilde{{\bm G}}$, and ${\bm B}$ 
are $2 \times 2$ matrices, and 
\be
\vec{\mathcal{X}}^{t}=\left(\zeta, 
\delta \rho_{m}/k \right) \,.
\ee
For sufficiently small-scale modes deep inside 
the sound horizon, we split $\tilde{{\bm G}}$ into the form 
\be
\tilde{{\bm G}}=\frac{k^2}{a^2}{\bm G}+{\bm M}\,,
\ee
where the leading-order terms of ${\bm G}$, 
${\bm M}$, and ${\bm B}$ are of order $k^0$.
The nonvanishing components of ${\bm K}$ 
and ${\bm G}$ are the following diagonal 
ones \cite{Kase:2018aps}, 
\ba
&&
K_{11}=\frac{\dot{\phi}^2 q_t q_s}{(2Hq_t-\dot{\phi} D_6)^2}\,,
\qquad 
K_{22}=\frac{a^2}{2(\rho_m+P_m)}\,,
\nonumber \\
&&
G_{11}=-q_tc_t^2-\frac{\rho_m+P_m}{2Hq_t
-\dot{\phi} D_6}{\cal F}_1
+\frac{1}{a}\frac{{\rm d}}{{\rm d}t}\left(a{\cal F}_1\right)\,,\nonumber \\
&&G_{22}=\frac{a^2c_m^2}{2(\rho_m+P_m)}\,,
\label{KGu}
\ea
where 
\be
{\cal F}_1=\frac{2q_t^2}{2Hq_t-\dot{\phi} D_6}\,.
\ee
The absence of off-diagonal components in 
${\bm K}$ and ${\bm G}$ implies that $\delta \rho_m$ 
is decoupled from $\zeta$ for sufficiently small-scale modes. 
In the perfect-fluid sector, the absence of ghost and Laplacian instabilities requires that $K_{22}>0$ 
and $G_{22}>0$, which translate to $\rho_m+P_m>0$ 
and $c_m^2>0$.
The no-ghost condition for $\zeta$ is given by 
\be
q_s^{(\rm u)} \equiv K_{11}
=\frac{\dot{\phi}^2 q_t q_s}{(2Hq_t-\dot{\phi}D_6)^2}>0\,. 
\label{qsue}
\ee
So long as $q_s>0$, along with the inequality $q_t>0$, 
the condition (\ref{qsue}) is satisfied.

The squared propagation speeds $c_{\rm s}^2$ 
for $\zeta$ and $\delta \rho_m$ can be obtained 
from the dispersion relation  
${\rm det} \left( c_{\rm s}^2 {\bm K}-{\bm G} \right)=0$. 
One of them is the value $c_m^2=G_{22}/K_{22}$ 
for $\delta \rho_m$, while another one 
for $\zeta$ is given by 
\be
c_s^2=\frac{G_{11}}{K_{11}}
=-\frac{c_t^2D_6^2+2B_1D_6+4q_t D_2}{q_s}\,, 
\label{cs}
\ee
where 
\ba
B_1 &=& 
\frac{2}{\dot{\phi}} \left[ \dot{q}_t+
(1-c_t^2) H q_t \right]\,,\\
D_2 &=&-\frac{1}{2}G_{2,X}
- 2 H\dot{\phi} G_{3,X} 
- H^2 (3 G_{4,X}+5\dot{\phi}^2 G_{4,XX}) 
 \nonumber \\
& &
-[2G_{4,X} + 2 \dot{\phi}^2 G_{4,XX} 
+ H \dot{\phi} ( 2 G_{5,X}+\dot{\phi}^2G_{5,XX})]
\dot{H} \nonumber \\
& &-\frac{1}{2} [ 2 G_{3,X}+\dot{\phi}^2 G_{3,XX}
+ 4 H \dot{\phi} ( 3 G_{4,XX}
\nonumber \\
& &
+\dot{\phi}^2 G_{4,XXX}) 
+ H^2 ( 2G_{5,X}+5\dot{\phi}^2 G_{5,XX}
\nonumber \\
& &
+\dot{\phi}^4 G_{5,XXX})]\ddot{\phi}-H^3\dot{\phi}
(2G_{5,X}+\dot{\phi}^2 G_{5,XX})\,.
\ea
To avoid the Laplacian instability for $\zeta$, we 
require that $c_s^2>0$.

\subsection{Model 1}

In Model 1, we begin by analyzing the stability of tensor and scalar perturbations along the solution $\dot{\phi}=H$, 
which arises when ${\cal C}=0$. 
We then extend our analysis to the case where
${\cal C} \neq 0$.

In the tensor sector, substituting the coupling functions 
(\ref{example1}) into Eqs.~(\ref{qt}) and (\ref{ct2}) gives 
\ba
q_t &=& 
\frac{1-\ell^2 \dot{\phi}\,(5\dot{\phi}-4H)}
{(1-\ell^2 \dot{\phi}^2)^3}\,,
\label{qtmo1}\\
c_t^2 &=&
\frac{1+\ell^2 (4\ddot{\phi}-\dot{\phi}^2)}
{1-\ell^2 \dot{\phi}\,(5\dot{\phi}-4H)}\,.
\label{ctmo1}
\ea
Along the solution $\dot{\phi}=H$, 
Eqs.~(\ref{qtmo1}) and (\ref{ctmo1}) reduce, 
respectively, to
\ba
q_t &=& \frac{1}{(1-\ell^2 H^2)^2}=\frac{(a^4+a_{m_1}^4)^2}{a^8}\,,
\label{qtmo1a}\\
c_t^2 &=& 1+\frac{4\ell^2 \dot{H}}{1-\ell^2 H^2}
=\frac{a^4-7a_{m_1}^4}{a^4+a_{m_1}^4}\,, 
\label{ctmo1a}
\ea
where Eq.~(\ref{Hsomodel1}) 
is used in the second equalities of Eqs.~(\ref{qtmo1a}) 
and (\ref{ctmo1a}).
Thus, we have $q_t>0$, and $q_t$ diverges as $a \to 0$. 
Since $c_t^2<0$ for $a<7^{1/4}a_{m_1}$, 
tensor perturbations undergo Laplacian instability 
during inflation. 
In particular, as $a \to 0$, we have $c_t^2 \to -7$, 
indicating that the rapid growth of tensor perturbations 
violates the homogeneity of the Universe soon after the onset of inflation\footnote{
We note that $c_t^2$ quickly approaches $1$ once $a$ exceeds $a_{m_1}$.  Thus, the deviation of $c_t^2$ 
from 1 becomes extremely small in the late 
Universe---for instance,
$|c_t^2-1|={\cal O}(10^{-40})$ at $a=10^{10}a_{m_1}$. 
This behavior is consistent with the observational bound on 
$c_t^2$ derived from the gravitational wave event GW170817 \cite{LIGOScientific:2017zic}.
However, the negative value of $c_t^2$ during inflation already spoils the successful cosmic expansion history.}.

In the scalar sector, the quantity (\ref{qsue}) is expressed as 
\begin{widetext}
\ba
q_s^{(\rm u)} &=&
6 \ell^2 \dot{\phi}^2 [1 - \ell^2 \dot{\phi} ( 5 \dot{\phi}-4H)] 
[1 -\ell^4 \dot{\phi}^2 (5 \dot{\phi}-2H)^2   
+5 \ell^6 \dot{\phi}^4 (5 \dot{\phi}^2- 6 H \dot{\phi}+2 H^2 ) 
+ \ell^2 (7 \dot{\phi}^2- 6 H \dot{\phi}+2 H^2)] 
(\dot{\phi}-H)^2\nonumber \\
& &
/\{ (1-\ell^2 \dot{\phi}^2)^3 [H-5H \ell^2 \dot{\phi}^2 
(2+3\ell^2 \dot{\phi}^2)+2\ell^2 \dot{\phi}^3
(1+5\ell^2 \dot{\phi}^2)+6H^2 \ell^2 \dot{\phi}
(1+\ell^2 \dot{\phi}^2)]^2 \}\,.
\label{qsu}
\ea
\end{widetext}
Along the solution $\dot{\phi}=H$, we have 
\be
q_s^{(\rm u)}=0\,,
\label{qsu}
\ee
which holds at all times during inflation and 
the subsequent cosmological epoch.
We note that the quantity $q_s$, which is defined 
in Eq.~(\ref{qsdef}), is also proportional to 
$(\dot{\phi}-H)^2$, and vanishes in the limit 
$\dot{\phi} \to H$. 
Since the numerators in Eqs.~(\ref{dotH}) and (\ref{ddotphi}) are both proportional to $(\dot{\phi}-H)^2$, $\dot{H}$ and 
$\ddot{\phi}$ remain finite even in the limit $\dot{\phi} \to H$, 
see Eq.~(\ref{atback4}).
However, the fact that $q_s^{(\rm u)}=0$ along the solution 
$\dot{\phi}=H$ implies that the kinetic term 
$q_s^{(\rm u)} \dot{\zeta}^2$ 
in the action (\ref{Ss2}) vanishes at all times.
This indicates a strong coupling problem, 
signaling a breakdown of perturbation theory due to the dominance of nonlinear over linear fluctuations.
From the onset of inflation, the strong coupling of nonlinear perturbations renders perturbative analysis on the homogeneous FLRW background invalid.

This issue does not stem from a particular gauge choice, 
but it arises in all physically meaningful gauges.
For example, we can choose the flat gauge $\zeta=0=E$ 
and consider $\delta \phi$ as a dynamical 
perturbation along with  $\delta \rho_m$. 
In this case, the kinetic term of the scalar 
field perturbation takes 
the form $q_s^{(\rm f)} \dot{\delta \phi}^2$, where 
$q_s^{(\rm f)}$ is related to $q_s^{(\rm u)}$
as $q_s^{(\rm f)}=(H^2/\dot{\phi}^2)
q_s^{(\rm u)}$ \cite{Kase:2018aps}.
Along the solution $\dot{\phi}=H$, we obtain 
$q_s^{(\rm f)}=q_s^{(\rm u)}=0$.
Thus, the strong coupling problem of the 
scalar field perturbation also manifests in the flat gauge. 
Since the properties $q_s^{(\rm u)}=q_s^{(\rm f)}=0$ hold at
all times along the solution $\dot{\phi}=H$, the use 
of the homogenous FLRW background is not legitimate throughout the cosmological evolution.

The denominator of $c_s^2$ is proportional to $q_s$, 
and thus it can be expressed as $b_0 (\dot{\phi}-H)^2$, where $b_0$ is a time-dependent coefficient. 
The numerator of $c_s^2$ can be written in the form  
$b_1 (\dot{\phi}-H)+b_2  (\ddot{\phi}-\dot{H})$, 
where $b_1$ and $b_2$ are time-dependent coefficients. 
Then, we can schematically express 
$c_s^2$ in the form 
\be
c_s^2=\frac{b_1 (\dot{\phi}-H)+b_2  (\ddot{\phi}-\dot{H})}
{b_0 (\dot{\phi}-H)^2}\,.
\label{csmo1}
\ee
Since we now consider the solutions $\dot{\phi}=H$ 
and $\ddot{\phi}=\dot{H}$, both the denominator and 
numerator of Eq.~(\ref{csmo1}) vanish. 
This implies that along the solution $\dot{\phi}=H$, 
which is realized for ${\cal C}=0$, the value of 
$c_s^2$ is generally undetermined.

When ${\cal C} \neq 0$, there is a deviation of $\dot{\phi}-H$ from 0. 
From Eqs.~(\ref{phiHre}) and (\ref{diffe}), we have 
$\ddot{\phi}=\dot{H} (1+\epsilon)+H \dot{\epsilon}$, 
where $\epsilon \simeq \epsilon_0 a^{13/3}$ and 
$\dot{\epsilon} \simeq 13H \epsilon/3$ in the regime 
$a \ll a_{m_1}$ (with $\epsilon_0$ being 
a constant). We substitute these solutions 
into the expressions of $q_t$ and $c_t^2$ and
use the leading-order solution of $H$ given 
in Eq.~(\ref{Hsomodel1}). Performing the expansion 
around $a=0$, it follows that 
\ba
q_t &=& 
\frac{a_{m_1}^8}{a^8}+{\cal O}(a^{-22/3})\,,\\
c_t^2 &=& -7+{\cal O}(a^{1/3})\,,
\ea
whose leading-order terms are equivalent to those 
derived for ${\cal C}=0$, see 
Eqs.~(\ref{qtmo1a}) and (\ref{ctmo1a}).
Thus, even when ${\cal C} \neq 0$, tensor perturbations still 
suffer from Laplacian instability during inflation.

Substituting $\dot{\phi}=H (1+\epsilon)$ as well as
its time derivative into $q_s^{({\rm u})}$ and $c_s^2$, 
and expanding them around $a=0$, we find
\ba
q_s^{(\rm u)} &=&
\frac{36 \epsilon_0^2 a_{m_1}^{16}}{a^{22/3}}
+{\cal O}(a^{-7})\,,\\
c_s^2 &=&
\frac{20}{27 \epsilon_0 a_{m_1}^4 a^{1/3}}
+{\cal O}(a^0)\,.
\ea
Thus, for $\epsilon \neq 0$, the leading-order term 
of $q_s^{(\rm u)}$ is nonvanishing and positive.
In the limit $a \to 0$, $q_s^{(\rm u)}$ diverges 
in proportion to $q_s^{(\rm u)} \propto a^{-22/3}$.
When $\epsilon_0<0$, i.e., ${\cal C}<0$, the scalar perturbation is subject to Laplacian instability  
due to a negative sound speed squared $c_s^2$.

For $\epsilon_0>0$, i.e., ${\cal C}>0$, 
$c_s^2$ can be positive during inflation, but it diverges 
as $c_s^2 \propto a^{-1/3} \to \infty$ when $a \to 0$. 
Let us choose a flat gauge and introduce a canonically normalized field $v=z \delta \phi$ in Fourier space, 
where $z=a \sqrt{2q_s^{(\rm f)}}$. 
For sufficiently small-scale modes where the kinetic and Laplacian terms dominate the second-order action 
of scalar perturbations, we obtain the following equation 
\be
v''+\left( c_s^2 k^2-\frac{z''}{z} \right) v=0\,,
\ee
where a prime denotes a derivative with respect to 
conformal time $\tau=\int a^{-1} {\rm d}t$. 
In the regime $a \ll a_{m_1}$, the quantity $z''/z$ can 
be estimated as $z''/z \simeq 40(aH)^2/9$. 
For the modes $c_s^2 k^2 \gg (aH)^2$, we may 
adopt the Bunch-Davies vacuum state, 
$v \simeq e^{-i c_s k \tau}/\sqrt{2c_s k}$. 
Then, the kinetic energy density of $\delta \phi$ 
can be estimated as $q_s^{(\rm f)}
|\dot{\delta \phi}|^2 \simeq c_s k/(4a^4)$, 
which diverges in the limit $a \to 0$. 
The same property also holds for the other energy density 
of $\delta \phi$ associated with the Laplacian 
term\footnote{In terms of the rescaled field 
$\tilde{\delta \phi}=k^{3/2}\delta \phi$, which has the dimension of mass,
it follows that 
$q_s^{(\rm f)}|\dot{\tilde{\delta \phi}}|^2 \simeq 
q_s^{(\rm f)} c_s^2 (k^2/a^2)|\tilde{\delta \phi}|^2
\simeq c_s k^4/(4a^4)$.}, such that 
$q_s^{(\rm f)} c_s^2 (k^2/a^2)|\delta \phi|^2 \simeq 
c_s k/(4a^4)$.
Despite the finiteness of the background field derivative 
$\dot{\phi}$, the energy density of linear 
scalar field perturbations diverges as  
$a \to 0$, indicating a breakdown of 
the perturbative treatment.

\subsection{Model 2}

In Model 2, the quantities relevant to the linear 
stability of tensor perturbations are given by 
\ba
q_t &=& 
\frac{1-\ell^4 \dot{\phi}^3 (5\dot{\phi}-4H)}
{(1-\ell^4 \dot{\phi}^4)^2}\,,\label{qtmo0}\\
c_t^2 &=&
\frac{1+\ell^4 \dot{\phi}^2  (4\ddot{\phi}-\dot{\phi}^2)}
{1-\ell^4 \dot{\phi}^3 (5\dot{\phi}-4H)}\,.
\label{qtmo2}
\ea
Along the solution $\dot{\phi}=H$, 
which can be realized for 
${\cal C}=0$, Eqs.~(\ref{qtmo0}) and (\ref{qtmo2}) 
reduce, respectively, to 
\ba
\hspace{-0.5cm}
q_t &=& 
\frac{1}{1-\ell^4 H^4}=\cosh \left( \frac{a_{m_2}^4}
{2a^4} \right)^2 \,,\label{qtmo0D}\\
\hspace{-0.5cm}
c_t^2 &=& 1+\frac{4\ell^4 H^2 \dot{H}}{1-\ell^4 H^4}
=1-\frac{4a_{m_2}^4}{a^4} 
\tanh \left( \frac{a_{m_2}^4}{2a^4} \right)\,,
\label{qtmo2D}
\ea
where we used Eq.~(\ref{Heq}).
The parameter $q_t$ remains positive and 
diverges for $a \to 0$.
In the regime $a \ll a_{m_2}$, one finds
$c_t^2 \simeq -4a_{m_2}^4/a^4$,  
diverging to $c_t^2 \to -\infty$ in the limit $a \to 0$.
This indicates a severe Laplacian instability of tensor perturbations, thus rendering the homogeneous 
background unstable.

The quantity associated with the no-ghost condition 
for scalar perturbations is given by 
\begin{widetext}
\ba
q_s^{(\rm u)} &=&
6 \ell^4 \dot{\phi}^3 [1 - l^4 \dot{\phi}^3 (5 \dot{\phi}-4H)]
[10 \ell^4 \dot{\phi}^3 H^2  - 2(1+15 \ell^4 \dot{\phi}^4)H
+ 25 \ell^4 \dot{\phi}^5 + 5 \dot{\phi}](\dot{\phi}-H)^2 
\nonumber \\
& &/[H + 2 \ell^4 \dot{\phi}^3 (5+3 \ell^4 \dot{\phi}^4) H^2 
+ 2 \ell^4 \dot{\phi}^5 (3+5 \ell^4 \dot{\phi}^4) 
- 3 \ell^4 \dot{\phi}^4 (6+5 \ell^4 \dot{\phi}^4)H]^2\,.
\label{qsumo2}
\ea
\end{widetext}
Since $q_s^{(\rm u)}=0$ along the solution $\dot{\phi}=H$,  
the strong coupling problem persists, as in Model 1.
The squared propagation speed for $\zeta$ takes the same form as in Eq.~(\ref{csmo1}), but with coefficients $b_0$, $b_1$, and $b_2$ that differ from those in Model 1.
Along the solution $\dot{\phi}=H$ 
(and hence $\ddot{\phi}=\dot{H}$), $c_s^2$ remains undetermined.

For ${\cal C} \neq 0$, we substitite $\dot{\phi}=H [1+\epsilon(t)]$ 
and its time derivative into the expressions of 
$q_t$, $c_t^2$, $q_s^{({\rm u})}$, and $c_s^2$, 
where $\epsilon(t) \simeq \epsilon_0 a^{-1} 
\exp(-a_{m_2}^4/a^4)$ in the regime $a \ll a_{m_2}$.
By using the solution (\ref{Heq}) and performing the 
expansion around $a=0$, it follows that 
\ba
q_t &=& a \exp (a_{m_2}^4/a^4) 
\left[ -\frac{1}{2\epsilon_0}+{\cal O}(a) \right]\,,\\
c_t^2 &=& -\frac{2a_{m_2}^4}{a^4}+{\cal O}(a^{-3})\,,\\ 
q_s^{(\rm u)} &=& a \exp (a_{m_2}^4/a^4) 
\left[ -\frac{1}{6\epsilon_0}+{\cal O}(a) \right]\,,\\
c_s^2 &=& -\frac{2a_{m_2}^4}{a^4}+{\cal O}(a^{-3})\,.
\ea
Since the leading-order terms of $c_t^2$ and $c_s^2$ 
are negative, both tensor and scalar perturbations 
undergo Laplacian instabilities during inflation. 
Moreover, both $c_t^2$ and $c_s^2$ diverge to 
$-\infty$ for $a \to 0$. 
When $\epsilon_0>0$ (i.e., ${\cal C}>0$), we find 
$q_t<0$ and $q_s^{(\rm u)}<0$ at leading order, 
implying ghost instabilities in both the tensor 
and scalar sectors.
When $\epsilon_0<0$, ghost instabilities are avoided, but 
Laplacian instabilities of tensor and scalar perturbations 
render the inflationary background illegitimate.

In Appendix, we also examine the linear stability of the background inflationary solution in Model 3. 
Again, the solution $\dot{\phi}=H$ suffers from the strong coupling problem in the scalar sector, as well as 
from a Laplacian instability in the tensor sector.
For ${\cal C} \neq 0$, Laplacian instabilities of both tensor and scalar perturbations arise during inflation, similar to Model 2 discussed above.

\subsection{Models of each power $n$}

In both Models 1 and 2, we have demonstrated the existence of a strong coupling problem for 
the solution $\dot{\phi}=H$.
This issue persists because the same behavior occurs 
for each power $n~(\geq 2)$ in the models discussed 
in Sec.~\ref{modeln}.
In these models, the quantity $q_s^{(\rm u)}$ 
takes the form 
\begin{widetext}
\ba
q_s^{(\rm u)} &=&
c_n \ell^{2n-2} n(n - 1)(2n - 3)
\dot{\phi}^{2n-3} \{ \ell^2 \dot{\phi}^3 + c_n \ell^{2n} 
n \dot{\phi}^{2n} [(3-2n)\dot{\phi} + 2(n - 1)H] \}
\{ l^2\dot{\phi}^3 [2(2 - n)H + (2n-1)\dot{\phi} ] 
\nonumber \\
& &
+ c_n \ell^{2n} n\dot{\phi}^{2n} 
[ (n - 1)(2n-1)H^2 - 2(n-2)(2n-1) H\dot{\phi} 
+ (n-2)(2n - 3) \dot{\phi}^2] \} (\dot{\phi}-H)^2
\nonumber \\
& &
/(\ell^2 H \dot{\phi}^3 + c_n \ell^{2n} n \dot{\phi}^{2n}
\{ (n - 1)(2n-1)H^2 -[3 + 4n(n-2)]H \dot{\phi} 
+ (n - 1)(2n - 3)\dot{\phi} ^2 \} )^2\,.
\ea
\end{widetext}
Along the solution $\dot{\phi}=H$, one finds 
$q_s^{(\rm u)}=0$ at all times.  
The property $q_s^{(\rm u)} \propto (\dot{\phi}-H)^2$ 
is preserved in models involving an infinite sum over $n$. 
This is why the strong coupling problem associated with
the background solution $\dot{\phi}=H$
persists for arbitrary coefficients $c_n$, 
even when the infinite sum is taken. 
The squared scalar sound speed takes 
the same form as Eq.~(\ref{csmo1}), so 
$c_s^2$ remains undertermined when 
$\dot{\phi}=H$ and $\ddot{\phi}=\dot{H}$.

Along the solution $\dot{\phi}=H$, the squared 
tensor propagation speed is given by 
\be
c_t^2=1+2(n-1)\frac{\dot{H}}{H^2} 
\left(1-\frac{1}{q_t} \right)\,,
\label{qtcnm}
\ee
where 
\be
q_t=1+c_n n (\ell H)^{2(n-1)}\,.
\ee
As long as $c_n>0$, the tensor ghost is absent. 
In the low-energy regime characterized by 
$c_n n (\ell H)^{2(n-1)} \ll 1$, one finds $c_t^2 \simeq 1$. 
By taking the infinite sum over $n$ like in Models 1 and 2, we have already seen that inflation 
can be realized with the divergence of $q_t$ as $a \to 0$. 
Since $\dot{H}<0$ during inflation, and the coefficient 
multiplying $\dot{H}/H^2$ in Eq.~(\ref{qtcnm}) 
can become large and 
positive when the infinite sum is taken, $c_t^2$ 
may turn negative. 
Indeed, we have seen that negative values of 
$c_t^2$ arise during inflation in all of Models 1, 2, 
and 3.

We note that 4DEGB gravity corresponds to $n=2$. 
In this case, we have 
\be
q_s^{(\rm u)}=\alpha_2 (\dot{\phi}-H)^2\,,
\ee
where 
\be
\alpha_2=\frac{6c_2 \ell^2 \dot{\phi}^2(1+2 c_2 \ell^2 H^2 )  
(1+4c_2 \ell^2 H \dot{\phi} - 2c_2 \ell^2 \dot{\phi}^2 )
}{(H+2c_2 \ell^2\dot{\phi}^3
+6c_2 \ell^2 H^2 \dot{\phi}-6 c_2 \ell^2 H \dot{\phi}^2)^2}.
\ee
The background scalar field equation is given by  
\be
4 c_2 \ell^2 \left( \dot{\phi}-H \right)^3
=\frac{{\cal C}}{a^3}\,.
\ee
For ${\cal C}=0$, the solution is uniquely determined as 
$\dot{\phi}=H$, and hence $q_s^{(\rm u)}=0$. 
If ${\cal C} \neq 0$, then $\dot{\phi}-H$ is nonzero 
and evolves as $\dot{\phi}-H \propto a^{-1}$. 
In the regime $c_2 \ell^2 H^2 \ll 1$, 
$q_s^{(\rm u)}$ approaches 0 on an expanding 
FLRW background.
Thus, as first pointed out in Ref.~\cite{Kobayashi:2020wqy}, 
the strong coupling problem is present for scalar cosmological perturbations in 4DEGB gravity.

%%%%%%%%%%%%%%%%%%%%%%%%%%%%%
\section{Conclusions} 
\label{consec}
%%%%%%%%%%%%%%%%%%%%%%%%%%%%%

In this paper, we have studied cosmology in four-dimensional theories arising from an infinite sum of 
Lovelock curvature invariants.
This is a generalization of 4DEGB gravity, in which the contribution of the higher-dimensional GB term is obtained through a rescaling of the coupling constant.
Under a conformal rescaling of the metric 
$\tilde{g}_{\mu \nu}=e^{-2\phi} g_{\mu \nu}$, 
the four-dimensional Lagrangian is obtained via 
dimensional regularization of the form (\ref{Ln}), 
which is evaluated for each value of $n=2,3,\cdots$.
Taking the infinite sum over $n$, the resulting action falls within a subclass of shift-symmetric Horndeski theories. 
The total Horndeski functions $G_{2,3,4,5}(X)$ are determined by specifying the coefficients $c_n$ 
in Eqs.~(\ref{G2})-(\ref{G5}).
In the main part of the paper, we considered two different choices for $c_n$, namely $c_n=1$ (Model~1) and 
$c_n=[1-(-1)^n]/(2n)$ (Model~2). 
In Appendix, we further analyzed the case 
$c_n=1/n$ (Model~3).

In Sec.~\ref{inssec}, we studied the background cosmological dynamics and showed that the solution 
for the scalar field current $J$ is given by $J={\cal C}/a^3$.
Since $J$ is proportional to $(\dot{\phi}-H)^3$ for each value of $n$, this property is preserved even when taking 
the infinite sum over $n$.
For ${\cal C}=0$, there exists a solution with 
$\dot{\phi}=H$, as confirmed in Models 1, 2, and 3. 
In the presence of a matter fluid such as radiation, 
this solution can give rise to singularity-free inflation with $H$ bounded from above, even in the asymptotic past.
In Models 1 and 2, the Hubble parameter can be 
expressed as Eqs.~(\ref{Hsomodel1}) and (\ref{Heq}), respectively, indicating that the inflationary period with a nearly constant expansion rate $H \simeq \ell^{-1}$ is
followed by the radiation-dominated era characterized by
$H \simeq \sqrt{\rho_{mi}}/(\sqrt{3}a^2)$. 
For ${\cal C} \neq 0$, the difference between 
$\dot{\phi}$ and $H$ is quantified by a time-dependent parameter $\epsilon(t)$ appearing in Eq.~(\ref{phiHre}).
In Models 1, 2, and 3, we estimated $|\epsilon(t)|$ by
Eqs.~(\ref{diffe}), (\ref{Hphi2}), and 
(\ref{epmo3}), respectively, in the regime $|\epsilon(t)| \ll 1$. In all these cases, $|\epsilon(t)|$ 
approaches 0 in the limit $a \to 0$. 

In Sec.~\ref{stosec}, we studied the consistency of 
cosmological solutions in regularized Lovelock gravity 
by applying the linear stability conditions for Horndeski theories previously derived in the literature. 
A common feature is that, along 
the solution $\dot{\phi}=H$, the kinetic term of the 
scalar field perturbation vanishes at all times 
(i.e., $q_s^{(\rm u)}=0$). 
This strong coupling problem arises not only in models 
with individual powers of $n$, but also persists in models constructed by taking the infinite sum over $n$. 
In other words, this problem cannot be circumvented 
by choosing specific values for the coefficients $c_n$. 
Unlike in some Horndeski genesis scenarios 
where strong coupling occurs only in the asymptotic past \cite{Ageeva:2020gti,Ageeva:2020buc}, scalar perturbations 
along the solution $\dot{\phi}=H$ are infinitely strongly coupled throughout the cosmological evolution. 
Moreover, we have shown that tensor perturbations exhibit Laplacian instabilities during inflation in all of 
Models 1, 2, and 3.

For ${\cal C} \neq 0$, i.e., when $\dot{\phi}$ slightly 
deviates from $H$, $q_s^{(\rm u)}$ does not vanish. However, both $q_s^{(\rm u)}$ and $c_s^2$ diverge in the limit $a \to 0$, with negative values of $c_s^2$ 
in Models 2 and 3. Despite the finite background 
kinetic term $\dot{\phi}^2$, the energy density of 
the scalar field perturbation diverges at the 
onset of inflation. 
Moreover, even when ${\cal C} \neq 0$, we found that $c_t^2$ is negative during inflation in all of Models 1, 2, 
and 3. These results indicate that, both for ${\cal C}=0$ and ${\cal C} \neq 0$, the use of a homogeneous cosmological background is not justified due to the dominance of inhomogeneities in the Universe.

We have thus shown that the four-dimensional theory obtained from the infinite sum of Lagrangians given in Eq.~(\ref{Ln}) does not yield a stable and physically viable inflationary solution that replaces the Big Bang singularity. 
It would be interesting to investigate whether spherically symmetric and static BHs or planar BHs in 
four dimensions \cite{Fernandes:2025fnz}, which may 
also exist in the same theory, exhibit similar pathologies. 
In 4DEGB gravity, it is known that strong coupling and instability issues arise for spherically symmetric BH solutions 
that respect asymptotic flatness \cite{Tsujikawa:2022lww}. 
Since the present four-dimensional theory belongs to a subclass of Horndeski theories, it should be straightforward to analyze the behavior of perturbations by using 
the linear stability conditions for spherically symmetric BHs derived in full Horndeski theories \cite{Kobayashi:2012kh,Kobayashi:2014wsa,Kase:2021mix, Kase:2023kvq}. 

There are several ways to avoid the strong coupling problem in 4DEGB gravity and its extensions. One approach is to break four-dimensional diffeomorphism invariance and construct theories that preserve only spatial diffeomorphisms within the Arnowitt-Deser-Misner (ADM) 
framework \cite{Aoki:2020lig}. 
Since the dynamical scalar degree of freedom is absent in such a scenario, constructing nonsingular cosmological or BH solutions may be challenging, even when 
an infinite sum of curvature invariants is considered.
However, when applied to cosmology and BH physics by including additional propagating degrees of freedom, such a theory should give rise to distinctive observational signatures---for instance, modifications to 
the non-Gaussianities of inflationary 
gravitational waves \cite{Aoki:2020ila}.

Another approach to avoiding the strong coupling problem is to introduce a four-dimensional action of the form 
$-\bar{M}^2 \int {\rm d}^4 x \sqrt{-\bar{g}}\,\bar{R}$ 
in addition to the Einstein-Hilbert action 
\cite{Gabadadze:2023quw}, 
where barred quantities are defined with respect to the conformally rescaled metric 
$\bar{g}_{\mu \nu}=e^{-2 \phi} g_{\mu \nu}$.
Below a cutoff scale $\bar{M}$, this prescription opens up 
the possibility of avoiding the strong coupling problem that arises in GR with trace anomalies \cite{Riegert:1984kt}.
Indeed, it was shown in Ref.~\cite{Tsujikawa:2023egy} that spherically symmetric and static BH solutions in the new gravitational theory with trace anomalies suffer from neither strong coupling nor instability problems. 
While this is an effective field theory valid up to the scale 
$\bar{M}$, it may be of interest to extend the analysis 
to the higher-dimensional spacetime and to investigate the implications of taking an infinite sum of curvature corrections. These issues are left for future work.

%%%%%%%%%%%%%%%%
\section*{Acknowledgements}
%%%%%%%%%%%%%%%%

The author was supported by JSPS KAKENHI 
Grant Number 22K03642 and Waseda University 
Special Research Project No.~2025C-488.

\section*{Appendix:~Model 3}
\renewcommand{\theequation}{A.\arabic{equation}} 
\setcounter{equation}{0}
\label{Appen}

In this Appendix, we address the linear stability 
of cosmological solutions for a model given by 
\be
c_n=\frac{1}{n}\qquad \quad ({\rm Model}~3)\,.
\ee
In this case, the Horndeski functions take 
the following form 
\ba
\hspace{-0.6cm}
G_2(X) &=& -\Lambda-\frac{2X (3-10\ell^2 X)}
{(1-2\ell^2 X)^2}-\frac{3}{\ell^2} 
\ln (1-2\ell^2 X)\,,  \nonumber \\
\hspace{-0.6cm}
G_3(X) &=& -\frac{2\ell^2 X (1+2\ell^2 X)}
{(1-2\ell^2 X)^2}\,, \nonumber \\
\hspace{-0.6cm}
G_4(X) &=& \frac{1}{2(1-2\ell^2 X)}\,,\nonumber \\
\hspace{-0.6cm}
G_5(X) &=& -\frac{\ell^2}{1-2\ell^2 X}+2\ell^2 
\tanh^{-1} \left( 1-4\ell^2 X \right)\,.
\label{example3}
\ea
The background equation for the scalar field 
is expressed as
\be
J=\frac{2 \ell^2 (1+3\ell^2 \dot{\phi}^2)}
{(1-\ell^2 \dot{\phi}^2)^3} (\dot{\phi}-H)^3
=\frac{{\cal C}}{a^3}\,.
\label{Jmo3}
\ee
For ${\cal C}=0$, there exists a solution $\dot{\phi}=H$. 
Setting $\Lambda=0$ and including radiation in the matter 
sector (with energy density $\rho_m=\rho_{mi}a^{-4}$), 
the background equations of motion along the solution 
$\dot{\phi}=H$ are given by 
\be
H=\frac{1}{\ell} \sqrt{1-\exp(-a_{m_1}^4/a^4)}\,,
\label{Hso3}
\ee
where $a_{m_1}=(\ell^2 \rho_{mi}/3)^{1/4}$, 
and 
\be
\dot{H}=\ddot{\phi}=-\frac{2}{3}(1-\ell^2 H^2) \rho_m\,.
\ee
In the limit $a \to 0$, $H$ approaches a constant value $\ell^{-1}$. 
The inflationary period, which occurs in the regime 
$a \ll a_{m_1}$, is followed by the radiation-dominated 
era for $a \gtrsim a_{m_1}$.

When ${\cal C} \neq 0$, the difference between $\dot{\phi}$ and $H$ can be quantified by the expression given 
in Eq.~(\ref{phiHre}).
Expanding in small values of $\epsilon(t)$, we obtain
\be
|\epsilon(t)| \simeq \left( \frac{|{\cal C}|}{2\ell^2 (1+3\ell^2 H^2)} 
\right)^{1/3} \frac{\exp(-a_{m_1}^4/a^4)}{aH}\,,
\label{epmo3}
\ee
where we used Eq.~(\ref{Hso3}).  
In the limit $a \to 0$, $|\epsilon(t)|$ approaches 0.
To keep the condition $|\epsilon(t)| \ll 1$ during inflation, 
we require that $|{\cal C}| \ll a_{m_1}^3/\ell$.

The quantities relevant to the linear stability of 
tensor perturbations are given by
\ba
q_t &=& 
\frac{1-\ell^2 \dot{\phi}\,(3\dot{\phi}-2H)}
{(1-\ell^2 \dot{\phi}^2)^2}\,,\label{qtmo3}\\
c_t^2 &=&
\frac{1+\ell^2 (2\ddot{\phi}-\dot{\phi}^2)}
{1-\ell^2 \dot{\phi}\,(3\dot{\phi}-2H)}\,.
\label{ctmo3}
\ea
Along the solution $\dot{\phi}=H$, we find
\be
q_t=\exp(a_{m_1}^4/a^4)\,,\qquad
c_t^2=\frac{a^4-a_{m_1}^4}{a^4}\,,
\ee
both of which diverge in the limit $a \to 0$. 
In particular, during inflation ($a \ll a_{m_1}$), 
$c_t^2 \simeq -a_{m_1}^4/a^4$, and hence tensor 
perturbations are subject to Laplacian instabilities.
Since $q_s^{(\rm u)}$ is proportional to 
$(\dot{\phi}-H)^2$, we have 
\be
q_s^{(\rm u)}=0\,,
\ee
along the solution $\dot{\phi}=H$. 
The squared scalar propagation speed is expressed 
in the form (\ref{csmo1}), so that $c_s^2$ 
is undetermined for $\dot{\phi}=H$.

For ${\cal C} \neq 0$, we consider the solution 
$\dot{\phi}=H[1+\epsilon(t)]$, where 
$\epsilon(t)=\epsilon_0 a^{-1}\exp(-a_{m_1}^4/a^4)$ 
in the regime $a \ll a_{m_1}$. 
Using the leading-order solution for $H$ given 
in Eq.~(\ref{Hso3}) and performing the expansion 
around $a=0$, it follows that 
\ba
q_t &=& a \exp (a_{m_1}^4/a^4) 
\left[ -\frac{1}{\epsilon_0}+{\cal O}(a) \right]\,,\\
c_t^2 &=& -\frac{2a_{m_1}^4}{a^4}+{\cal O}(a^{-3})\,,\\ 
q_s^{(\rm u)} &=& a \exp (a_{m_1}^4/a^4) 
\left[ -\frac{1}{3\epsilon_0}+{\cal O}(a) \right]\,,\\
c_s^2 &=& -\frac{2a_{m_1}^4}{a^4}+{\cal O}(a^{-3})\,.
\ea
Thus, both tensor and scalar perturbations exhibit 
Laplacian instabilities during inflation. 
Ghosts are present for $\epsilon_0>0$ and 
absent for $\epsilon_0<0$.
These properties are analogous to those 
observed in Model 2.

\bibliographystyle{mybibstyle}
\bibliography{bib}

\end{document}